%% file: main.tex
\documentclass[10pt,a4paper,oneside]{article}

\usepackage[utf8]{inputenc}
\usepackage{tabularx}
\usepackage{graphicx}
\usepackage{booktabs}
\usepackage{multirow}
\usepackage{amsmath}
\usepackage{enumitem}
\usepackage{url} 
\usepackage{tikz} 
\usepackage{varwidth} 
\usepackage{hyperref} 
\usepackage{color}
\usepackage{smartdiagram}
\usepackage{tikzscale}
\usepackage{todonotes} 
\usetikzlibrary{positioning,arrows,shapes.misc,arrows,shapes,shadows,trees,calc,fit,matrix,chains,intersections} 
\usepackage{lscape}
\usepackage{longtable}
\usepackage{adjustbox}
\usepackage{pifont}
\usepackage{authblk}

\tikzset{test/.style={font=\footnotesize}}

\newcommand{\bigyes}{\ding{52}\ding{52}}
\newcommand{\yes}{\ding{51}}
\newcommand{\no}{\ding{55}}
\providecommand{\keywords}[1]{\textbf{\textit{Index terms---}} #1}

\tikzset{
	basic/.style  = {draw, text width=2cm, drop shadow, font=\sffamily, rectangle},
	root/.style   = {basic, rounded corners=2pt, thin, align=center,
		fill=green!30},
	level 2/.style = {basic, rounded corners=6pt, thin,align=center, fill=green!60,
		text width=8em},
	level 3/.style = {basic, thin, align=left, fill=pink!60, text width=6.5em}
}

\begin{document}
	
	\title{The relationship between trust in AI and trustworthy machine learning technologies}
	
	\author[1]{Ehsan Toreini\thanks{ehsan.toreini@ncl.ac.uk}}
	\author[2]{Mhairi Aitken\thanks{mhairi.aitken@newcastle.ac.uk}}
	\author[2]{Kovila Coopamootoo\thanks{kovila.coopamootoo@newcastle.ac.uk}}
	\author[1]{Karen Elliott\thanks{karen.elliott@newcastle.ac.uk}}
	\author[1]{Carlos Gonzalez Zelaya\thanks{c.v.gonzalez-zelaya2@ncl.ac.uk}}
	\author[1]{Aad van Moorsel\thanks{aad.vanmoorsel@newcastle.ac.uk}}
	\affil[1]{School of Computing, Newcastle University, United Kingdom}
	\affil[2]{Business School, Newcastle University, United Kingdom}
	
	\date{}
	\maketitle
	
	\begin{abstract}
		To build AI-based systems that users and the public can justifiably trust one needs to understand how machine learning technologies impact trust put in these services. To guide technology developments, this paper provides a systematic approach to relate social science concepts of trust with the technologies used in AI-based services and products. We conceive trust as discussed in the ABI (Ability, Benevolence, Integrity) framework and use a recently proposed mapping of ABI on qualities of technologies. We consider four categories of machine learning technologies, namely these for Fairness, Explainability, Auditability and Safety (FEAS) and discuss if and how these possess the required qualities. Trust can be impacted throughout the life cycle of AI-based systems, and we introduce the concept of Chain of Trust to discuss technological needs for trust in different stages of the life cycle. FEAS has obvious relations with known frameworks and therefore we relate FEAS to a variety of international Principled AI policy and technology frameworks that have emerged in recent years. 
		
	\end{abstract}
	
	
	\section{Introduction}
	\label{s:intro}
	With growing interest in ethical dimensions of AI and machine learning has come a focus on ways of ensuring trustworthiness of current and future practices (e.g. European Commission 2019, IBM n.d.). The current emphasis on this area reflects recognition that maintaining trust in AI may be critical for ensuring acceptance and successful adoption of AI-driven services and products \cite{trustMLSocial,mayer1995}. This has implications for the many AI-based services and products that are increasingly entering the market. How trust is established, maintained or eroded depends on a number of factors including an individual's or group's interaction with others, data, environments, services, products and factors, which combine to shape an individual's perception of trustworthiness or otherwise. Perceptions of trustworthiness impact on AI and consequently, influence a person's decision and behaviour associated with the service or product. In this paper, we research the connection between trust and machine learning technologies in a systematic manner. The aim is to identify how technologies impact and relate to trust, and, specifically, identify trust-enabling machine learning technologies.  AI and machine learning approaches are {\em trustworthy} if they have properties that one is {\em justified} to place trust in them (see \cite{avizienis04} for this manner of phrasing).
	
	In the context of the FAT* conference, it is important to highlight the difference between studying trust in AI and studying ethics of AI (and data science). Trustworthy AI is related to normative statements on the qualities of the technology and typically necessitates ethical approaches, while trust is a response to the technologies developed or the processes through which they were developed (and may not necessarily - or entirely - depend on ethical considerations). Ethical considerations behind the design or deployment of an AI-based product or service can impact perceptions of trust, for instance if trust depends on having confidence in the service not discriminating against the trusting entity (or in general). However, there may be cases where ethics is not a consideration for the trusting entity when placing trust in a service, or, more frequently, if ethics is one of the many concerns the trusting entity has in mind. In what follows, we see that trust-enhancing machine learning technologies can be related to Principled AI frameworks (such as Asilomar AI Principles introduced in 2017, Montr\'eal Declaration for Responsible Development of Artificial Intelligence in 2018 and IEEE Ethically Aligned Design Document), in addition, we consider a wider set of technologies than typically follow on from considering the technology implementation of Principled AI frameworks.   
	
	\begin{figure}[htbp]
		\includegraphics[width=0.97\textwidth]{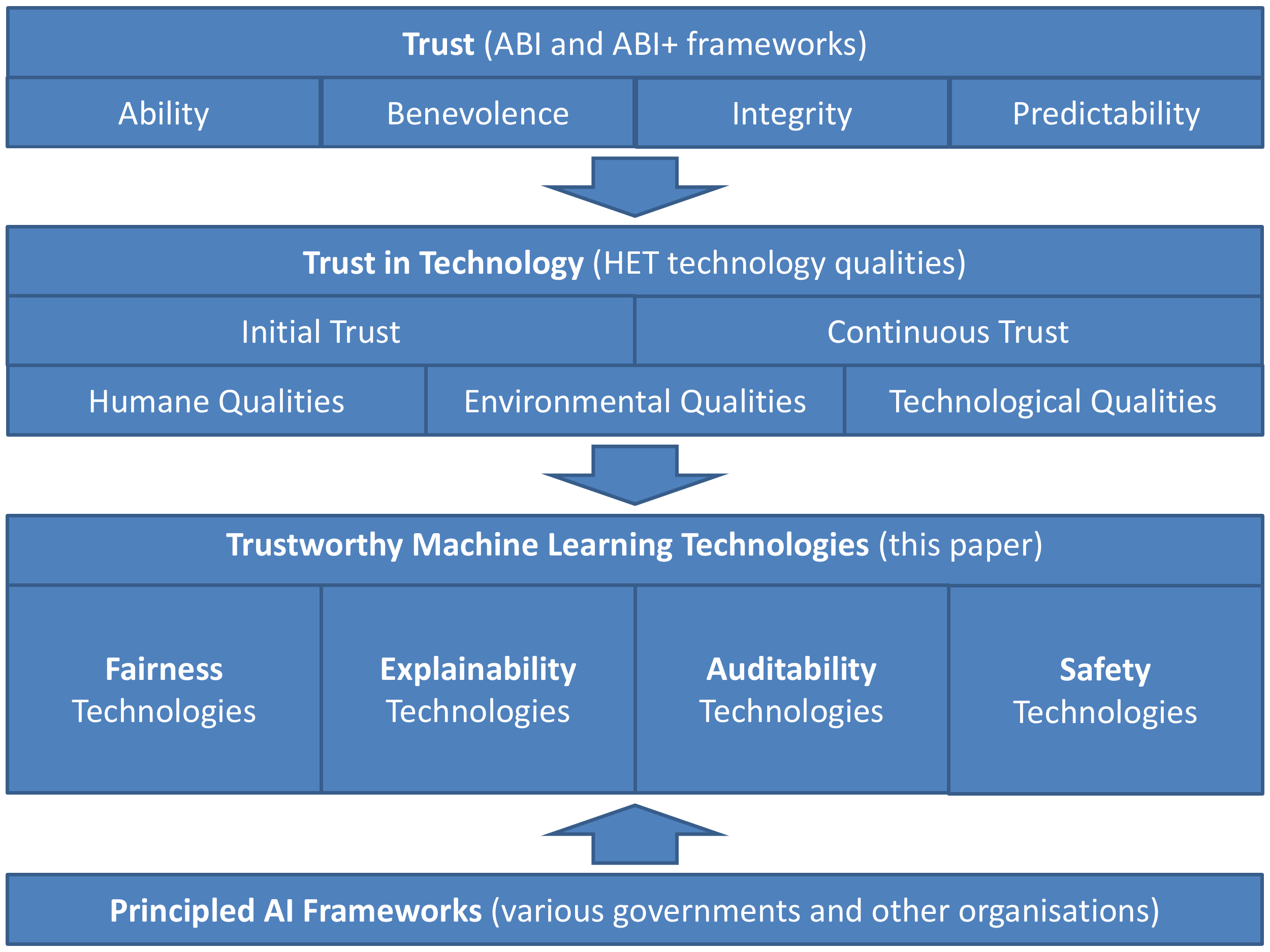}
		\caption{The contribution of this paper: identification of trust-enhancing Machine Learning technologies based on social sciences literature and relating these with Principled AI frameworks}
		\label{fig:outline}
	\end{figure}
	
	The aim of this paper is to identify the ways in which (classes of) machine learning technologies might enhance or impact trust, based on trust frameworks drawn from the social sciences and trust in technology literature. Figure \ref{fig:outline} outlines our approach.  At the centre of Figure \ref{fig:outline} is the end product of this paper, trust-enhancing technologies in machine learning and their classification in technologies for Fairness, Explainability, Auditability and Safety (FEAS).  The downwards arrows indicate that these technologies are derived from trust frameworks from social science literature (particularly organisational science).  The upward arrow indicates that the FEAS-classification of technologies was also informed from the various Principled AI frameworks that shape the ethics and policy discussion in many nations (this is discussed in Section \ref{ss:FAT*}). 
	
	As indicated in Figure~\ref{fig:outline}, we base our discussion on the widely accepted ABI (Ability, Benevolence, Integrity) principles underlying trust, as introduced by Mayer et al.\cite{mayer1995} and extended to include the predictability element by Dietz and Den Hartog \cite{dietz2006}.  We add to this a temporal dimension, from initial trust to continuous trust, as discussed by Siau et al.\cite{socialTrustTechnology}. This gives us a base to understand trust in general, and we augment this further by integrating Siau`s perspective on trust in technology, which identifies that trust is impacted by Human, Environmental and Technological qualities (referred to as the technologies' HET qualities in what follows).  We will discuss these steps to go from the ABI+ model to HET qualities of trustworthy technologies in Section \ref{s:trust}. 
	
	To summarise, the contributions of this paper are as follows:
	\begin{itemize}
		\item We draw on social science literature, particularly from organisational science, to apply established principles of trust to examine the qualities for technologies to support trust in AI-based systems (primarily based on the ABI and ABI+ framework and the HET qualities). 
		\item We identify how trust can be enhanced in the various stages of an AI-based system's life space and through the design, development and deployment phases of a system life-cycle. We therefore introduce the concept of an AI \emph{Chain of Trust} to discuss the various stages and their interrelations.
		\item We introduce a FEAS (Fairness, Explainability, Auditability, Safety) classification of machine learning technologies that support and enable trust and establish the relation between these trust-enhancing technologies and the HET qualities.
		\item We discuss how our technology classification and trustworthy machine learning techniques relate to various Principled AI framework considered by policy makers and researchers in ethics and associated topics.
	\end{itemize}
	
	\section{Trust}
	\label{s:trust}
	This section discusses trust frameworks we can use to classify and identify trustworthy machine learning technologies. It discusses the top half of the paper contribution provided in Figure \ref{fig:outline}, the box with Trust and with Trust in Technology. In Section \ref{ss:ABI} we introduce the ABI+ framework to describe trust in general, i.e., not restricted to trust in technology. Section \ref{ss:technology} reflects on trust in technology and science, recognising that AI-based services are based on science and manifest as technologies. The final sections provide the framework developed by Siau, which includes a discussion on time-sensitivity of trust (Section \ref{ss:time_trust}) and recognises three types of qualities technologies may exhibit that impact trust (Section \ref{ss:siau}).
	
	Trust is discussed across many diverse social science literature leading to an abundance of definitions and frameworks available through which to examine the concept. It is a concept which in everyday conversation is routinely and intuitively used and yet remains challenging to define and study. Andras et al.~\cite{andras2018} summarise some of the ways that trust has been approached across different disciplines: ``In the social world trust is about the expectation of cooperative, supportive, and non-hostile behaviour. In psychological terms, trust is the result of cognitive learning from experiences of trusting behaviour with others. Philosophically, trust is the taking of risk on the basis of a moral relationship between individuals. In the context of economics and international relations, trust is based on calculated incentives for alternative behaviours, conceptualised through game theory.' A comprehensive review of literature relating to trust is beyond the scope of this paper. Here we focus on established models to examine the nature of trust and discuss how this relates to technology.
	
	
	\subsection{The ABI Framework: Ability, Benevolence and Integrity}
	\label{ss:ABI}
	The ABI framework introduced by Mayer et al~\cite{mayer1995} suggests that the three main attributes which will shape an assessment of the trustworthiness of a party are: Ability, Benevolence and Integrity. The model discusses the interactional relationship between a trustor (the entity that trusts) and a trustee (the entity to be trusted).  Building on the work of Mayer et al~\cite{mayer1995}, Dietz and Den Hartog~\cite{dietz2006} outline three forms of trust: trust as a belief; a decision and; an action. While many studies have focused on trust as a belief in isolation from actions, Dietz and Den Hartog~\cite{dietz2006} regard the three forms as being the constituent parts of trust which are most usefully examined together. 
	
	In the ABI model, ability is defined as the perception: ``that group of skills, competencies, and characteristics that enable a party to have influence within some specific domain''.  The \emph{specific domain} is crucial as assessments of a party`s ability will vary according to particular tasks or contexts. Benevolence is defined as ``the extent to which a trustee is believed to want to do good to the trustor''. To be considered to possess Integrity a trustee must be perceived to adhere ``to a set of principles that the trustor finds acceptable''. This requires confidence that the trustee will act in accordance to a set of principles and that those align with the values of the trustor. 
	
	Since its inception Mayer et al.~\cite{mayer1995}’s framework has been adapted and expanded to acknowledge the importance of Predictability or Reliability in shaping perceived trustworthiness.Dietz and Den Hartog~\cite{dietz2006} developed the ABI+ model suggesting that the four key characteristics on which judgements of trustworthiness are based are: Ability; Benevolence; Integrity and; Predictability. Predictability will reinforce perceptions of the Ability; Benevolence and; Integrity of the trustee. Considering the role of Predictability, draws attention to the importance of trust being sustained overtime through ongoing relationships. 
	
	While each of the attributes are related and may reinforce one another they are also separable~\cite{mayer1995}. One party may trust another even if they perceive one or more of these attributes to be lacking. As such trust -and trustworthiness- should not be thought of in binary terms but rather trust exists along a continuum. Dietz and Den Hartog~\cite[p.~563]{dietz2006} suggest five degrees of trust ranging in intensity. 
	1.	“deterrence-based trust” which they do not in fact consider to be trust but rather an example of distrust as it means that the trustor has confidence to take actions as the other party is limited in their own possible actions to such an extent that they do not perceive risk (e.g. due to regulation).
	2.	“calculus-based trust” which is considered “low trust” and still involves considerable suspicion.
	3.	“knowledge-based trust” considered as “confident trust”
	4.	“relational-based trust” considered as “strong trust”
	5.	“identification-based trust” considered as “complete trust”
	
	As such evaluating one party`s trust in another is more complex than a straightforward assessment of whether or not they trust that party and statements such as ``A trusts B'' are overly-simplistic, instead trust is described in statements reflecting the conditional nature of trust, for example: ``A trusts B to do X (or not to do Y), when Z pertains…''~\cite[p.~564]{dietz2006}.
	
	As well as being shaped by judgements of trustworthiness the act of trusting also depends on a range of external and contextual factors, personal attributes and traits of the \emph{trustor}. An individual`s predisposition or ideological position will impact on the extent to which they trust particular individuals/organisations, and these positions will shape how they receive, interpret and respond to information about the other party~\cite{dietz2006}. 
	
	As the context changes, for example relating to cultural, economic, political or personal developments, so levels of trust and perceptions of trustworthiness also change. Therefore trust is characterised as an ongoing relationship rather than a static concept. Moreover, trust can be strengthened – or conversely weakened – through interactions between trustors and trustees: ``outcomes of trusting behaviours will lead to updating of prior perceptions of the ability, benevolence, and integrity of the trustee''~\cite[p.~728]{mayer1995}. See also Section~\ref{ss:time_trust} for a discussion of the time- -sensitive nature of trust. 
	
	
	\subsection{Trust in Science and Technology}
	\label{ss:technology}
	AI-based services and products often have strong scientific element through the use of powerful algorithms in innovative ways, and a very pronounced technology flavour through the automated manner in which such services and products are used. There is a significant body of literature in the field of Science and Technology Studies~(STS) examining public relationships with science and technology, and, in particular, the role of trust in these relationships. In 2000, the UK House of Lords Science and Technology Committee published a landmark statement (which continues to be widely cited) stating that there was a ‘crisis of trust in science’. This reflected wider discourses suggesting that a series of high-profile scientific controversies and scandals (e.g. BSE, thalidomide and the MMR triple vaccine), together with the rapid pace of scientific progress had resulted in an erosion of public trust in science~\cite{aitken2016}. 
	
	This led to considerable attention directed at ‘improving’ public trust in science, typically through efforts to increase public understanding of science, on the assumption that, where the public is sceptical or mistrusting, this can be explained by ignorance or lack of understanding, and as such can be ‘corrected’ through better dissemination of scientific knowledge or \emph{facts}~\cite{aitken2016}). However, such approaches are now widely discredited as it is recognised that they overlook the role of members of the public in actively engaging with scientific knowledge rather than being “passive recipients of scientific knowledge”~\cite[p.~206]{cunningham2006}. Members of the public critically assess, deconstruct, question and evaluate claims to scientific knowledge in line with their own ideologies, experiences and the contexts in which the information is received~\cite{hagendijk2006}. This active process shapes people's trust as beliefs as well as informing the trust decisions and actions that are taken.
	
	
	The public`s relationship with science and technology is too sophisticated to be characterised by a simple trust/distrust binary relationship. Rather, in many cases the public adopts an ambivalent form of trust – described by Wynne~\cite{wynne2006,wynne1992misunderstood,wynne1996reflexive} as an: \emph{as if} trust. This takes account of the public's ``knowingly inevitable and relentlessly growing dependency upon expert institutions''~\cite[p.~212]{wynne2006}.  
	
	With regard to AI-based technologies, the dependence on knowledge and behaviours of experts is clear, and trust is increasingly conditional.  This implies that people do not automatically have confidence in particular innovations, scientists or scientific institutions (but equally lack of absolute trust does not mean that innovations will be met with public opposition). There can be dissonance between the trust beliefs held and the decisions and actions taken based on contextual, personal or organisational factors. In particular, even where people do not fully trust the technology they may use a service driven by AI if they feel there is no alternative option. 
	
	\subsection{Trust Technology Qualities: Humane, Environmental and Technological}
	\label{ss:siau}
	To further understand how technology interfaces with trust, Siau et al.~\cite{trustMLSocial} identify qualities of technologies that relate to trust and the concepts in the ABI+ framework of Section \ref{ss:ABI}.  The authors recognise three types of conditions to demonstrate the potential for a technology to be perceived as trustworthy: humane, environmental and technological qualities.
	
	\paragraph{Humane Qualities.\ } Humane qualities refer to the actions that attract individuals possessing a risk--taking attitude. The effectiveness of this quality depends on the personality type, past experiences and cultural backgrounds of the individuals. This is linked to the ability of the trustee to satisfy the curiosity of the trustor in testing a desired task. In other words, if cultural background resonates and if testing a product or service is feasible, this will typically enhance trust.
	
	\paragraph{Environmental Qualities.\ } Environmental qualities consider elements that are enforced by the qualities of the technology provider. First, it heavily relies on the nature of the task that the technology handles. The sophistication of the task has a potential to attract trustworthiness or cause distraction. The pattern of the establishment of trustworthiness differs in various places depending on the education system, level of their accessibility to novel modern advancements and subtle inherent cultural backgrounds. Yuki et al.~\cite{yuki2005cross} discussed such cultural impact on the trust establishment pattern. Institutional factors are another environmental parameter for trust. Siau et al.~\cite{trustMLSocial} defined it as ``the impersonal structures that enable one to act in anticipation of a successful future endeavor''. They collected two aspects for this concept: the institutional normality and structural assurance. The first deals with efficiency of the organisational structure and the later refers to the ability of an institution to fulfil the promises and its commitments.
	
	\paragraph{Technological Qualities.\ } Finally, technological qualities determine the capacity of the technology itself to deliver the outcome as promised. This commitment is multi-dimensional. First, the technology needs to yield the results efficiently. Thus, it needs to establish an agreed performance metric and assure its outcome yields into the desirable range of the metric. Second, the technology should define concrete boundaries for their solution. The user~(as potential trustors) of the technology should be provided with enough information to infer the purpose of the technology and set their expectations sensibly based on such understanding. Lastly, The process of the technology outcome is another factor in trustworthiness. The technology should be able to reply to potential queries about \emph{how} they concluded such outcome and \emph{why} it led to the such performance. This aspect outlines the relation between the performance and purpose aspects. 
	
	\subsection{Time-domain: Initial and Continuous Trust}
	\label{ss:time_trust}
	A final element to support our understanding of trustworthy technologies is understanding trust as it develops over time, as also discussed in \cite{socialTrustTechnology}. Highlighting the importance of predictability in the ABI+ model, the dynamics of relationship between trustor and trustee is an ongoing process. Usually, it requires initial preconditions to be satisfied, provided through first impressions, which is referred to as \emph{initial trust}. After the initial phase, trust levels may change, for a variety of reasons, and this is referred to as \emph{continuous trust} in the literature~\cite{trustMLSocial}. Typical examples that may impact continuous trust in AI-based services and products are data breaches, privacy leaks or news items on (unethical) business or other practices.
	
	In what follows we will consider an additional time-sensitive element for trustworthy AI-based technologies, namely that of the service and product life cycle, both in terms of moving between the stages of design, development and deployment, as well as in terms of the machine learning pipeline, which includes data input, algorithm selection and output presentation.  See the next section, and particularly Section \ref{ss:chain}. 
	
	\section{Machine Learning and the Chain of Trust}
	\label{s:ml}
	In this section we introduce the concept of a Chain of Trust, which connects trust considerations in the stages of the machine learning pipeline and, when considered over time, the AI-based service or product may iterate through these stages (effectively expending the chain into a cycle, as illustrated in Figure \ref{fig:pipeline}). Before introducing the Chain of Trust in Section \ref{ss:chain}, we briefly review the basics of machine learning, as well as the notion of a machine learning pipeline.  
	
	\subsection{Basics of Machine Learning}
	\label{ss:ml}
	A machine learning algorithm is basically a function as $y = f_{\theta}(x)$ (with exception of a few algorithms such as nearest--neighbour~\cite{altman1992introduction}). In this equation, $f_{\theta}$ represents the function that maps input to the output, i.e. the machine learning model. In this function, $\theta \in \Theta$ denotes a set of values that is tuned for the optimal operation of $f$. $\theta$ is calculated based on a pre-defined loss function that measures the similarity or dissimilarity between samples. The input of a model, $x$, correspond to a set of \emph{features} which is a vector of values that represents to data, a.k.a. dataset. Finally, $y$ represents the output of the algorithm to fulfil the task that it meant to undergo, i.e. \emph{supervised}, \emph{unsupervised} and \emph{reinforcement} learning. 
	
	In supervised learning, the output $y$ is meant to be an assignment of the input $x$ to a pre-defined label set.  Supervised Learning algorithms are mainly used in object recognition~\cite{krizhevsky2012imagenet}, machine translation~\cite{sutskever2014sequence}, filtering spams~\cite{drucker1999support}, etc. For example, a fraud detection classifier would assign two labels~(fraud or benign) to an input feature which is derived from a transaction.
	
	Unsupervised Learning methods are used when the input $x$ and output $y$ are both unlabelled. In these methods, the task is to determine a function $f_{\theta}$ that takes $x$ as input and detects a hidden pattern representation as output, $y$.  The problems that unsupervised learning tackles include grouped a dataset based on a similarity metric~(a.k.a clustering~\cite{jain1999data}), projecting data to a reduced dimension space~(e.g. PCA methods~\cite{krizhevsky2009learning}) and pre--training algorithms for the other tasks~(i.e. pre--processing methods)~\cite{erhan2010does}.
	
	Reinforcement Learning~\cite{sutton2018reinforcement} methods maps $x$ to a set of policies as $y$. In these techniques, $f_{\theta}$ determines an action, an observation or a reward~($y$) that should be taken into account when situation $x$ is observed. Reinforcement learning techniques are concerned with how an \emph{agent}, suppose a rescue robot, should behave under certain circumstances to maximise the chances of fulfilling a purpose.
	
	\subsection{Machine Learning Pipeline}
	\label{ss:pipeline}
\begin{figure}
		\centering	
		\includegraphics[width=0.85\textwidth,trim={5cm 3.5cm 5cm 2cm} ,clip]{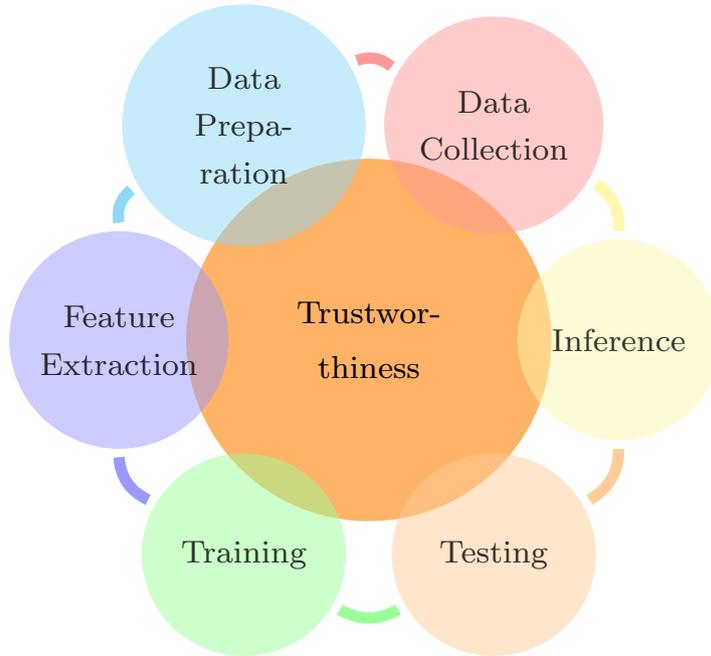}
		\caption{Various stages in the machine learning pipeline, motivating the notion of a Chain of Trust.}
		\label{fig:pipeline}
	\end{figure}

	Regardless of the task, the use of any machine learning algorithm implies activities in various stages, called the {\em pipeline}. This is depicted in Figure \ref{fig:pipeline} through the chain of circles.  First, one collects the data from a source and store the digital representation in a database (`Data Collection' in Figure \ref{fig:pipeline}) . Then, such data undergoes pre-processing methods to extract certain features, as $x$ in the machine learning equation, labelled `Data Preparation' and `Feature Extraction', respectively. 
	
	When $x$ is ready to be processed for the supposed task, the features are divided into at least two groups to attain two purposes. The first group of features (`Training' in Figure \ref{fig:pipeline}) are used to tune $\theta$ to optimise the output ($y$) of the function $f$. The group of features for this purpose is called the training data. The training process can be offline or online, depending on how static the training data is with respect to the life-cycle of a model. The online fashion deals with dynamic training in which the model re-tunes $\theta$ when new training data arrives. In contrast, in offline mode, the training stage only operates once on a static training dataset. 
	
	The second group of features is used for the purpose of verifying the generalisation of the $f_{\theta}$ parameters when the model faces an unknown parameters when new training data appears in the process (`Testing' in Figure \ref{fig:pipeline}). Verification is done by assessing the efficiency of the performance through some chosen metric. For example, in classification, a typical accuracy metric is the proportion of the test data that has been mapped to their original label correctly by $f_{\theta}$. The features and data used at the testing stage are called \emph{test data}. When the model passes the verification stage with a sufficiently good performance, then they are applied in the wild. This stage is known as `Inference', in which the trained model is deployed to face unseen data. In this context, $\theta$ is fixed and $y$ is computed for $f_{\theta}(x)$ when $x$ is unknown~(i.e. not contained in either test and training datasets).  
	
	In what follows we group some of the stages in the machine learning pipeline. The first group of stages concerns data-centric aspects, involving data collection methods, pre-processing techniques and extraction of useful features for the analysis. The second group of stages is model-centric, the stages in Figure \ref{fig:pipeline} that deal with the tuning the model to the best performance (`Training' stage), evaluating the trained model for confirmation of the desirable performance (`Testing' stage) and deployment of the model for the real-world application (`Inference' stage). 
	
	\subsection{Chain of Trust}
	\label{ss:chain}
	We are now in a position to introduce the notion of Chain of Trust. Based on the machine learning pipeline depicted in Figure \ref{fig:pipeline} it becomes clear that technologies may impact trust in the resulting service or product in various stages of the pipeline.  For instance, better methods to clean the data during `Data Preparation' may avoid bias in the output of algorithms, which in turn  helps to enhance trust once it becomes visible to users or the public.  There are a number of important dimensions to the Chain of Trust, each of which demonstrates the importance of continuous trust as discussed in Section \ref{ss:time_trust}.  
	
	First, stages may impact on each other in terms of the level of trust they are able to propagate. Romei et al.~\cite{interdisciplinary} reviewed various cases studies that led to biased decisions and analysed the causes. There is an important specific case of this, namely that trust impact may only manifest itself in later stages or at a later time.  For instance, in the above example of improving the `Data Preparation', this enhancement will only impact trust by users if it is made visible to these users.  This may for instance be through better results when using the services, but, possibly more likely, may also only become visible if news article or long-term statistics are presented to users that explain that results are, say, less biased and that therefore can be trusted. However, where trust is established or damaged based on visible outcomes at later stages the resulting levels of trust will have implications for trust in all stages of the development of future technologies.   
	
	The second dimension present in the Chain of Trust results from the fact that a service or product may iterate through the stages during its lifecycle, possibly multiple times.  This occurs, for instance, when new data is being introduced to improve the `Training', and through this, the `Inference'.  In this case, effectively the service cycles through the chain depicted in Figure \ref{fig:pipeline}.  
	
	A third dimension within the notion of Chain of Trust is the development of trust through the stages of design, development and deployment of the AI-based service or product. Trust will be impacted by technology decisions in all stages of the lifecycle. In the above examples, we mainly consider the deployment stage, in which trust is considered from the perspective of a running service or existing product. Even simply making an AI-based service available may run the risk of introducing new biases, or exacerbating existing ones, because changing the way a service is offered may imply it is less useful or effective for certain groups. For instance, a recent study demonstrates inherent discrimination in automated mortgage advising in the FinTech industry~\cite{fintechFairness}.  Therefore, to establish trusted AI-based solutions it will be critical to consider trust from the initial stages, starting from the design of a new service or product.  In so doing, trust is considered {\em a priori}, before it is being deployed, and does not come as a surprise once the service is running. 
	
	A final dimension to consider in the context of the Chain of Trust is that of accidents, sudden breakdowns of trust or failures.  Typical examples of such trust failures are security breaches that impact trust, (reports about) data loss of the service or similar services, or the discovery of bias in the machine learning results that drive a service.  Such accidents can take place through all pipeline and life cycle stages discussed above and may have severe impact on the level of trust users place in AI-based systems. However, the ways in which an organisation responds to such accidents can be equally, or more, important for determining the impact on trust. Dietz and Gillespie \cite{dietz2012} have shown that scandals or crises which risk damaging the reputation of an organisation can also act as catalysts for culture change bringing about and reinforcing new ethical/trustworthy practice. They are opportunities to forge new relationships with stakeholders (positive or negative) \cite{gillespie2009}. Technology solutions that continuously monitor and possibly transparently share data about service bias are trustworthy technologies that may assist in avoiding or mitigating the impact of trust failures.  
	
	\section{Trust in AI-Based Systems}
	\label{s:trustAI}
	We now aim to establish the connection between {\em trust} in the AI-based solution and {\em trustworthiness} of the underlying technologies.  We first discuss in Section \ref{ss:trustconcerns} various trust related issues one may encounter in AI-based services and products.  We then discuss in Section \ref{ss:FAT*} a variety of existing Principled AI policy and technology frameworks that have emerged in recent years. Based on these frameworks, we will propose in Section \ref{ss:FEAS} a technology-inspired Principled AI variant, namely FEAS Technologies, that is, technologies for Fairness, Explainability, Auditability and Safety.  
	
	\subsection{Trust Considerations in Stages of the Chain of Trust}
	\label{ss:trustconcerns}
	In this section we discuss trustworthiness aspects in the various stages identified in the Chain of Trust. We divide the discussion in data-related concerns (with the focus on data collection, data pre--processing and feature selection, Section \ref{ss:data-specific}) and model-related concerns (with the focus on the stages of model training, testing, and inference, Section \ref{ss:model}).  
	
	\subsubsection{Data Related Trust Concerns: Data Collection and Pre-Processing Stages.\ }
	\label{ss:data-specific}
	Data collection involves reading data from various sources reliably (e.g. sensors to collect environmental data, a smart speaker listening to audio commands or website that stores session-related data from the user's browser, or etc.); secure transmission of the data from the collection point to a machine for the purpose of storage or online analysis; storing data in server. The pre--processing stage includes mechanisms to clean the dataset by removing null, duplicate or noisy data. Usually, pre--processing solutions combines with the other ones in data--related stages.
	
	One of the main trust concerns in data is in protection against privacy violations and other forms of illegitimate use of one's data. Legislative support (e.g, in Europe via GDPR regulation) has an important role to play.  GDPR is a perfect illustration of several data-related trust issues. For instance, the collection process should be transparent and requires explicit consent from users for many data uses. Moreover, the subject keeps the right to challenge the ability to collect data, has control over their collected data and maintains the ``right to be forgotten''. GDPR-type concerns and solutions provides a basis to identify technologies that enhance trust.   
	
	In general, the perceived intrusive nature of instances of data collection has convinced common users to be more cautious in the situations that their data is being collected~\cite{chin2012measuring}. This is a threat to trustworthiness of a service that functions based on collected data, particularly AI-based services and products.  
	
	Technological solutions for trustworthy data collection, pre-processing as well as storage have been well established.  Trustworthiness usually relies on factual declaration of the good faith (benevolence) and abiding to it in action (integrity). The current trust challenges are therefore less in the development of technology solutions than in identifying ways of interacting, working and regulating aspects that impact trust.  
	
	\subsubsection{Model-Related Trust Concerns: Feature Extraction, Training, Testing and Inference Stages\ }
	\label{ss:model} 
	Model-related trust concerns trust in the working of the models and algorithms.  This set of concerns gets to the heart of trust challenges we phase in a world in which AI becomes omni-present, fundamentally changing the way people live their lives.   Many of the trust concerns relate to FAT, a fear that algorithms may harm society, or individuals, because they are unfair, non-transparent and without accountability.  As for data (Section \ref{ss:data-specific}) GDPR is useful as an illustration of the issues important for society. GDPR enforces a requirement to ``explain'' results of AI-based solutions to end users, which implies a desire for more transparent machine learning. GDPR also contains substantial accountability measures to implement a mechanism to ``challenge'' the outcome of AI--based technology.
	
	The question is what the role of technologies in dealing with model-related trust concerns.  At their core, machine learning models and algorithms are optimised for accuracy of results and efficiency in obtaining these. While the accuracy and speed of results might be considered to demonstrate basic functionality (ability in ABI+ terminology, Section \ref{ss:ABI}), they do not necessarily satisfy or align with other trust qualities. In most cases, algorithms are considered as a ``black-box'' \cite{blackBoxMachineLearning}, which implies algorithms can be assessed only in relation to the outcomes they produce. Assessments of the benevolence or integrity (again using ABI+ terms) can therefore only be done in indirect manners, through that of the entity that develops or applies the AI-based solution. Of course, justifying such trust is problematic, especially in the time of high profile data breaches and scandals relating to mishandling or misuse of personal data~(e.g. Facebook, Cambridge Analytica). 
	
	In Section \ref{s:FEAS} we will introduce a number of technologies to enhance or impact trust, of two types. The first type is to establish a mechanism to verify the outcomes of the model. In this case, an agent would be responsible to function in parallel to the model and the model outcome are not accessible until the agent and the model are both satisfied in a pre--defined criteria. In the second type one endeavours to (re)design a model or choice of algorithms into something that is inherently more trustworthy. For example, the design of a fair SVM model refers to embedding a fairness constraints into its definition so that the model functions with the built-in consideration of that notion.  This set of approaches we will discuss in detail in Section \ref{s:FEAS}.

	\subsection{Principled AI Policy Frameworks}
	\label{ss:FAT*}
	The trust concerns discussed in the previous section are related to concerns that have been raised widely about the impact on society of the proliferation of AI.  This has resulted in the emergence of a large amount of policy frameworks that relate to Principled AI frameworks, that is, policy frameworks to enhance and regulate Fairness, Accountability and Transparency of, particularly, AI-based services and products.  Principled AI frameworks are particularly relevant to trust as well. Therefore, as depicted in Figure \ref{fig:outline}, the bottom box, it is opportune to relate Principled AI frameworks to the trustworthy technology classification we introduce in Section \ref{ss:FEAS}.  It is important to note that in this paper we use FEAS to classify {\em technologies}, while many of the Principled AI frameworks inform {\em policy} and do not provide much detail in terms of specifying or restricting technology implementations to achieve the policy objectives.  
	
	Principled AI frameworks have been introduced by various stakeholders~(technology companies, professional, standardisation, governmental and legislator bodies, academic researchers), as illustrated by Table \ref{tbl:frameworks}. These Principled AI frameworks present varying sets of qualities that AI-based systems should follow to be considered trustworthy (some Principled AI frameworks (also) apply to technologies other than AI).  In general, these documents present high--level definitions for the objectives and qualities of the involved science and technology, but do not go into the specifics of technical implementation guidelines. 
	
	There is emerging literature reviewing Principled AI frameworks. Whittlestone et al.~\cite{whittlestone2019role} provide a critical analysis of frameworks for ethical machine learning and highlights a number of challenges, some of which require attention from a technical perspective.  For instance, frameworks may confuse the interpretation of qualities, present conflicting definitions and/or qualities may be different across different documents. Particularly relevant also for the underlying technologies is that the frameworks often fail to realise dependencies between policy objectives (e.g. addressing discrimination issues might lead to unexpected privacy leakages). Current frameworks are focused on privacy, transparency and fairness issues but this needs to be shifted toward understanding such tensions and re-framing core research questions. 
	
	In yet unpublished work, Fjeld et al.~\cite{ethicalGraph} analyse currently available Principled AI frameworks, from industry, governments, general public, civil societies and academic bodies.  Table~\ref{tbl:frameworks} contains many of the Principled AI frameworks considered by \cite{ethicalGraph} (see Section \ref{ss:FEAS} for an explanation of how we compiled Table \ref{tbl:frameworks}). Interestingly, they recognised 47 qualities, categorised them into eight groups, which are a combination of the qualities identified by Siau~\cite{socialTrustTechnology}, i.e., humane (promotion of human values, professional responsibility, human control of technology) and technological qualities (fairness and non--discrimination, transparency and explainability, safety and security, accountability, privacy). The authors have available a graphical demonstration of their findings ~\footnote{\url{https://ai-hr.cyber.harvard.edu/primp-viz.html}}. 
	
	We note again that most of the frameworks do not focus on trust, but on ethics, privacy and related concerns. Moreover, the terms \emph{ethical} and \emph{trustworthy} machine learning are at times used interchangeably in these frameworks \cite{koszegi2019high}. This would effectively imply that trustworthiness is achieved through abiding with the ethical concepts such as human rights or non-discrimination approaches. However, while ethical considerations are inevitably related to perceptions of trust, ethical machine learning and trustworthy machine learning are not necessarily the same thing.  Specifically, in terms of ABI+, ethical machine learning would necessarily emphasise the benevolence aspects of trust, while the other two aspects critical for trust (ability and integrity) are insufficiently represented.
	
	\begin{landscape}
		\scriptsize{
			\input{frameworks.tex}
		}
	\end{landscape}
	
	\subsection{FEAS: Fair, Explainable, Auditable and Safe Technologies}
	\label{ss:FEAS}
	We propose to classify trustworthiness technologies in Fair, Explainable, Auditable and Safe Technologies (FEAS).  This is in part motivated by a desire to align our discussion of trustworthy technologies with the Principled AI frameworks that are available in the literature, as discussed in Section \ref{ss:FAT*}.  We converged on FEAS based on our knowledge and understanding of the technologies involved, classified in manner we believe will be comprehensible, illustrative and natural for technologists.  Note that the FEAS technological qualities are in addition to the essential technological qualities of \emph{accuracy} and \emph{efficiency} and performance of the algorithm(s), without which trustworthiness is not possible. 
	
	\begin{itemize}
		\item {\bf Fairness Technologies:\ } technologies focused on detection or prevention of discrimination and bias in different demographics \cite{fairnessOverview,statiticalFramework,interdisciplinary,disparateOdds2,counterfactual,corbett2018measure,Joseph2016a,demographic,equalizedOdds,dataScienceView,fairclassifier,zafar2019discrimination,fairnessAwareness}.
		\item {\bf Explainability Technologies:\ } technologies focused on explaining and interpreting the outcome to the stakeholders~(including end-users) in a humane manner~\cite{athey2015machine,lipton2016mythos,Blum1997,Goncalves2010,Uysal2014,duwairi2014study,Cai2018,zelaya2019volatility,rivest1987learning,Lou2013,wood2003thin,montavon2018methods}.
		\item {\bf Auditability Technologies:\ } technologies focused on enabling third-parties and regulators to supervise, challenge or monitor the operation of the model(s) \cite{lineage1,lineage2,lineage3,lineage4,lineage5,lineage11,lineage12,lineage6,lineage7,lineage8,lineage9,lineage10}.
		\item {\bf Safety Technologies:\ } technologies focused on ensuring the operation of the model as intended in presence of active or passive malicious attacker~\cite{securityassessment,MLSecuritySurvey,bishop2007penetration,biggio2011support,trainingAttack2,trainingAttack6,MPCCrypto,secureCompuation}. 
	\end{itemize}
	
	\subsubsection{FEAS Related to Principled AI.\ } 
	Table \ref{tbl:frameworks} provides the relationship between the FEAS technology classes and the Principled AI frameworks identified in \cite{ethicalGraph}. We reviewed each of the frameworks with respect to the FEAS technology classes required to establish the qualities mentioned in the framework. We mark frameworks that refer to fairness, explainability, safety and auditability qualities using the symbols explained in the caption of Table \ref{tbl:frameworks}. 
	
	As one sees immediately from Table \ref{tbl:frameworks} all frameworks are related to FEAS technologies, and would be able to make use, or even require, FEAS technologies to be available. Note that there is a considerable difference in the granularity of the discussions in the Principled AI frameworks compared to that of the computing literature.  Hence, the precise technology needs for each framework would need deeper investigations, and may not be completely specified within the existing framework documents.  For instance, the policy frameworks refer to the general existence of discrimination caused by bias in machine learning algorithms, but in the technological literature there are at least 21 mathematical definitions for fairness and a wide range of solutions to prevent/detect bias in the algorithm.  The technology discussion in Section \ref{s:FEAS} is therefore at a much deeper level of detail than that of the Principled AI frameworks of Table \ref{tbl:frameworks}.  
	
	\begin{table}[]
		\caption{Trustworthy technology classes versus trust qualities \cite{socialTrustTechnology}. \no = \emph{less important}, \yes = \emph{important}, \bigyes\emph{very important}}
		\begin{tabular}{@{}l|c|c|c|c@{}}
			\toprule
			& Fair & Explainable  & Auditable  & Safe \\ \midrule
			Humane Qualities   & \bigyes       & \bigyes            & \yes          & \no     \\ \midrule
			\begin{tabular}[c]{@{}l@{}}Technological Qualities\end{tabular} & \yes      & \yes            & \no          & \bigyes    \\ \midrule
			\begin{tabular}[c]{@{}l@{}}Environmental Qualities\end{tabular}  & \bigyes      & \bigyes            & \yes          & \yes    \\ \bottomrule
		\end{tabular}
		\label{tbl:exampleQualities}
	\end{table}
	
	\subsubsection{FEAS Related to Trust Qualities.\ }
	Table \ref{tbl:exampleQualities} provides the relationship between trust qualities (humane, environmental and technological, see Section \ref{ss:siau}) and FEAS technologies.  Table \ref{tbl:exampleQualities} is based on the authors' understanding of the qualities and technologies, the latter to be discussed deeper in Section \ref{s:FEAS}.  Fairness strongly requires technologies that are strong in humane and environmental qualities, as does explainability, since their effectiveness strongly depends on individuals and the culture or setting.  Safety is dominated by technological quality associated with security and reliability of the systems.  
	
	
	\section{Trustworthy Machine Learning Technologies}
	\label{s:FEAS}
	This section discuss technologies for trustworthy machine learning using the FEAS grouping introduced in the previous section.  We introduce the FEAS classes, discuss challenges and provide some examples of existing approaches.  A full review of technologies is beyond the scope of this paper.  
	
	\subsection{Fairness Technologies}
	This group of technologies is concerned about achieving fair, non-discriminating outcomes. The ethical aspects of fairness constitute a structural assurance in the service or product that enhances (or at least impacts) trust. 
	
	Fair machine learning is a difficult challenge. A first challenge is to identify if how to measure unfairness, typically in terms of bias or related notions. Narayanan \cite{narayanan2018translation} has identified at least 21 definitions of fairness in the literature, which cannot necessarily all be obtained at the same time.  To enhance trust, the metrics used by machine learning experts needs to relate to how it impacts trust by individuals and the public, posing an additional challenge.  
	
	Various solutions has been proposed to establish a subset of fairness notions. One approach is to reduce the bias in the dataset, known as \emph{debiasing data}. However, it is not sufficient (or even helpful) to simply ignore or remove features associated with unfairness, e.g. gender, ethnics ~\cite{statiticalFramework}.  Luong et al.~\cite{luong2011k} assigned a decision value to each data sample and adjusted this value to eliminate discrimination.  Kamishima et al.~\cite{fairclassifier} proposed a model--specific approach by adding a regulating term for a logistic regression classifier, which eventually leads to unbiased classification. Other fairness mitigation solutions focus on the inference stage. They enforce the output of the model to generate an specific notion of fairness~\cite{equalizedOdds}.  
	
	\v{Z}liobait{\.e} et al.~\cite{fairnessOverview} identify two conditions for non-discriminating machine learning: data with the same non-protected attributes should give the same outcomes, and the ability to distinguish outcomes should be of the same order as the difference in the non-protected attribute values. This provides a more generic understanding helpful to design fairness technologies. This paper does not aim to review all techniques but it is clear that many challenges remain in achieving fair machine learning technologies. 
	
	\subsection{Explainability Technologies}
	Explainability refers to relating the operation and outcomes of the model into understandable terms to a human~\cite{kim2015ibcm}. In the machine learning literature, notions of \emph{explainability}, \emph{transparency}, \emph{intelligibility},\emph{comprehensibility} and \emph{interpretability} are often used interchangeably~\cite{explainableSurvey}. Lipton~\cite{lipton2016mythos} provides a thorough discussion on these terms and the differences. He also concludes that explainability increases trust.
	Doshi-Velez et al.~\cite{doshi2017towards} suggested two types of explainability, namely explainability of an application (e.g. a physician being able to understand the reasons for a classifier's medical diagnostic~\cite{lime}) and explainability to understand the way in which a classifier is coming up with its outputs, mainly by using intelligible models.
	
	There are two main approaches in explainable solutions. The first one, known as \emph{ex-ante}, refers to the use of highly intelligible models to obtain the desired predictions. The second one, known as \emph{ex-post}, refers to the use of a second model to understand the learned model's behaviour. In ex--ante approach, an explicit prediction function analyses feature coefficients to understand their impact over a decision, decision trees or decision lists~\cite{rivest1987learning}). 
	
	Ex-post explanations are categorised into \emph{local} and \emph{global} explainers.
	The ex-post approach uses explain. There is a fast growing body of literature, including local explainers (e.g., \cite{lime,shrikumar2017learning} and the unified framework for local explainers which generalises these and other existing methods \cite{lundberg2017unified}), global explainers (e.g., \cite{Lakkaraju2017}). There are many challenges left with respect to explainable machine learning technologies, including trading off fidelity to the original model and explainability \cite{montavon2018methods}.
	
	
	
	\subsection{Auditability Technologies}
	Auditability technologies refer to methods that enable third parties to challenge the operation and outcome of a model. This provides more transparency to black--box characteristics of machine learning algorithms. Enabling machine learning lineage provides insights into how they have operate, which gives a greater degree of transparency compared to explainability of current processes or outcomes.
	
	More specifically, auditability in machine learning may involve assessing the influence of input data in the output of the model. This ensures predictability, a process that is also referred as ``decision provenance'' in the literature~\cite{singh2018decision}. Singh et al.~\cite{singh2018decision} argue that decision provenance have two objectives, it should provide the history of particular data and it should provide a viewpoint on the system`s behaviour and its interactions with its inner components or outside entities.
	
	Much of the literature around auditability relates to data provenance research in database and cloud concepts~\cite{lineageSurvey} and model provenance approaches~\cite{ghodsi2017safetynets}. It involves proposals on how to store provenance data~\cite{lineage1,lineage2,lineage3,lineage4,lineage5}, summarisation of provenance data~\cite{lineage6,lineage7,lineage8}, specific query language for the provenance data~\cite{lineage7}, query explainability~\cite{lineage9,lineage10}, Natural language processing for provenance data~\cite{lineage11,lineage12} and cryptographic solutions to verify model's behaviours without exposing user privacy~\cite{papernot2016towards} or revealing model`s intellectual properties~\cite{walfish2015verifying}.
	
	\subsection{Safety Technologies}
	Data and the machine learning model could both be the target to adversarial operations. The extent of attacker's access to the data and model depends on the intention of the attack and the weaknesses in the system`s architecture. The attacker can execute a targeted attack to harm an individual or perform a indiscriminate attack. Moreover, the attacker can perform the attack stealthily for the intention of information gathering or intelligence~(a.k.a exploratory attack) or he/she can actively engage into the functioning of the system for the purpose of manipulation~(a.k.a causative attack). 
	
	The security and privacy foundations of a ML model is not different from classical security model of Confidentiality, Integrity and Availability~(CIA model~\cite{MLSecuritySurvey}).  We omit availability, since its relevance is general, not just or specifically for AI-based services. Careless preparation of the stored data would leak information to an attacker~\cite{securitySurvey} but confidentiality can be enhanced in many ways, for instance through approaches for differential privacy \cite{MLSecurity3,dwork2011differential}, homomorphic encryption \cite{hu2017generating} or cryptography integrated in machine learning algorithms \cite{MPCCrypto}.  Integrity can be enhanced by either preventing tampering, such as in pre-processing \cite{MLSecuritySurvey}, or by discovering and possibly repairing tampered data \cite{securityassessment,quantitativeSecurity2,quantitativeSecurity1}. These methods deal only with data, but it is also relevant to consider integrity during the execution of the algorithm, e.g., in the training stage \cite{demontis2017yes,trainingAttack3,trainingAttack4,securitySurvey,szegedy2013intriguing}.
	
	\section{Conclusion}
	\label{s:conclusion}
	This paper established the connection between trust as a notion within the social sciences, and the set of technologies that are available for trustworthy machine learning. More specifically, we related the ABI framework and HET technology qualities for trust with categories of machine learning technologies that enhance trustworthiness. We identified four categories of technologies that need to be considered: Fair, Explainable, Auditable and Safe (FEAS) technologies. These need to be considered in various interrelated stages of a system life cycle, each stage forming part of a Chain of Trust. The paper shows a close relationship between technologies to improve the trustworthiness of AI-based systems and those that are being pursued in ethical AI and related endeavours. We illustrated this by mapping of FEAS technologies on concerns in a large set of international Principled AI policy and technology frameworks. 
	
	
	\bibliographystyle{plain}
	\bibliography{technical,social}
	
	
	
\end{document}

%% file: frameworks.tex
\begin{longtable}{ | p{5cm} | *{15}{c|} }
	\label{tbl:frameworks}\\
	\caption{Trustworthy technology classes related to FAT* frameworks. \no = \emph{no mention}, \yes = \emph{mentioned}, \bigyes = \emph{emphasised}}
	\\
	\hline
	\toprule
	Framework & Year & Document Owner & Entities & Country        & Fairness & Explainability & Safety & Auditability \\ 
	\hline
	\endhead  
	\midrule

	Top 10 principles of ethical AI                                                                         & 2017 & UNI Global Union                                                                                                                       & Ind                                                       & Switzerland    & \yes      & \yes            & \yes      & \yes          \\ \midrule
	Toronto Declaration                                                                                     & 2018 & Amnesty International                                                                                                                  & Gov, Ind                                                  & Canada         & \bigyes   & \yes            & \no       & \yes          \\ \midrule
	\begin{tabular}[c]{@{}l@{}}Future of work and Education\\ For the Digital Age\end{tabular}              & 2018 & T20: Think 20                                                                                                                          & Gov                                                       & Argentina      & \bigyes   & \yes            & \yes      & \yes          \\ \midrule
	Universal Guidelines for AI                                                                              & 2018 & The public voice coalition                                                                                                             & Ind                                                       & Belgium        & \bigyes   & \yes            & \yes      & \yes          \\ \midrule
	Human Rights in the Age of AI                                                                           & 2018 & Access Now                                                                                                                             & Gov, Ind                                                  & United States  & \bigyes   & \yes            & \bigyes   & \yes          \\ \midrule
	Preparing for the Future of AI                                                                          & 2016 & \begin{tabular}[c]{@{}l@{}}US national Science,\\ and Technology Council\end{tabular}                                                  & \begin{tabular}[c]{@{}l@{}}Gov, Ind, \\ Acad\end{tabular} & United States  & \yes      & \yes            & \bigyes   & \yes          \\ \midrule
	Draft AI R\&D Guidelines                                                                                 & 2017 & Japan Government                                                                                                                       & Gov                                                       & Japan          & \no       & \yes            & \bigyes   & \yes          \\ \midrule
	White Paper on AI Standardization                                                                     & 2018 & Standards Administration of China                                                                                                      & Gov, Ind                                                  & China          & \yes      & \no             & \bigyes   & \bigyes       \\ \midrule
	\begin{tabular}[c]{@{}l@{}}Statements on AI, Robotics and \\ `Autonomous' Systems\end{tabular}          & 2018 & \begin{tabular}[c]{@{}l@{}}European Group on Ethics in \\ Science and New Technologies\end{tabular}                                    & \begin{tabular}[c]{@{}l@{}}Gov, Ind,\\ Acad\end{tabular}  & Belgium        & \yes      & \yes            & \yes      & \bigyes       \\ \midrule
	\begin{tabular}[c]{@{}l@{}}For a Meaningful Artificial\\  Intelligence\end{tabular}                     & 2018 & \begin{tabular}[c]{@{}l@{}}Mission assigned by \\ the French Prime Minister\end{tabular}                                               & Gov, Ind                                                  & France         & \yes      & \yes            & \no       & \bigyes       \\ \midrule
	AI at the Service of Citizens                                                                           & 2018 & Agency for Digital Italy                                                                                                               & Gov, Ind                                                  & Italy          & \yes      & \yes            & \yes      & \yes          \\ \midrule
	AI for Europe                                                                                           & 2018 & European Commission                                                                                                                    & Gov, Ind                                                  & Belgium        & \yes      & \yes            & \yes      & \yes          \\ \midrule
	AI in the UK                                                                                            & 2018 & UK House of Lords                                                                                                                      & Gov, Ind                                                  & United Kingdom & \bigyes   & \yes            & \yes      & \yes          \\ \midrule
	AI in Mexico                                                                                            & 2018 & British Embassy in Mexico City                                                                                                         & Gov                                                       & Mexico         & \yes      & \no             & \yes      & \yes          \\ \midrule
	Artificial Intelligence Strategy                                                                        & 2018 & \begin{tabular}[c]{@{}l@{}}German Federal Ministries \\ of Education, Economic Affairs, \\ and Labour and Social Affairs\end{tabular} & Gov, Ind                                                  & Germany        & \yes      & \yes            & \yes      & \bigyes       \\ \midrule
	\begin{tabular}[c]{@{}l@{}}Draft Ethics Guidelines for\\  Trustworthy AI\end{tabular}                    & 2018 & \begin{tabular}[c]{@{}l@{}}European High Level Expert \\ Group on AI\end{tabular}                                                      & Gov, Ind, Civ                                             &                & \bigyes   & \yes            & \bigyes   & \yes          \\ \midrule
	AI Principles and Ethics                                                                                 & 2019 & Smart Dubai                                                                                                                            & Ind                                                       & UAE            & \bigyes   & \yes            & \bigyes   & \yes          \\ \midrule
	\begin{tabular}[c]{@{}l@{}}Principles to Promote FEAT AI in \\ the Financial Sector\end{tabular}        & 2019 & Monetary Authority of Singapore                                                                                                        & Gov, Ind                                                  & Singapore      & \yes      & \yes            & \no       & \yes          \\ \midrule
	Tenets                                                                                                  & 2016 & Partnership on AI                                                                                                                      & Gov, Ind, Acad                                            & United States  & \yes      & \yes            & \yes      & \yes          \\ \midrule
	Asilomar AI Principles                                                                                  & 2017 & Future of Life Institute                                                                                                               & Ind                                                       & United States  & \yes      & \yes            & \yes      & \yes          \\ \midrule
	The GNI Principles                                                                                      & 2017 & Global Network Initiative                                                                                                              & Gov, Ind                                                  & United States  & \no       & \yes            & \yes      & \yes          \\ \midrule
	Montreal Declaration                                                                                    & 2018 & University of Montreal                                                                                                                 & Gov, Ind, Civ                                             & Canada         & \bigyes   & \bigyes         & \bigyes   & \yes          \\ \midrule
	Ethically Aligned Design                                                                                & 2019 & IEEE                                                                                                                                   & Ind                                                       & United States  & \yes      & \yes            & \yes      & \bigyes       \\ \midrule
	Seeking Ground Rules for AI                                                                             & 2019 & New York Times                                                                                                                         & Ind, GeP                                                  & United States  & \yes      & \yes            & \yes      & \yes          \\ \midrule
	\begin{tabular}[c]{@{}l@{}}European Ethical Charter on the Use\\ of AI in Judicial Systems\end{tabular} & 2018 & Council of Europe: CEPEJ                                                                                                               & Gov                                                       & France         & \yes      & \yes            & \yes      & \yes          \\ \midrule
	AI Policy Principles                                                                                    & 2017 & ITI                                                                                                                                    & Gov, Ind                                                  & United States  & \yes      & \yes            & \bigyes   & \yes          \\ \midrule
	The Ethics of Code                                                                                      & 2017 & Sage                                                                                                                                   & Ind                                                       & United States  & \yes      & \no             & \no       & \yes          \\ \midrule
	Microsoft AI Principles                                                                                 & 2018 & Microsoft                                                                                                                              & Ind                                                       & United States  & \yes      & \yes            & \bigyes   & \yes          \\ \midrule
	AI at Google: Our Principles                                                                            & 2018 & Google                                                                                                                                 & Ind                                                       & United States  & \yes      & \yes            & \bigyes   & \yes          \\ \midrule
	AI Principles of Telef\textbackslash{}'onica                                                            & 2018 & Telef\textbackslash{}'onica                                                                                                            & Ind                                                       & Spain          & \yes      & \yes            & \yes      & \no           \\ \midrule
	\begin{tabular}[c]{@{}l@{}}Guiding Principles on Trusted \\ AI Ethics\end{tabular}                      & 2019 & Telia Company                                                                                                                          & Ind                                                       & Sweden         & \yes      & \yes            & \bigyes   & \yes          \\ \midrule
	\begin{tabular}[c]{@{}l@{}}Declaration of the Ethical \\ Principles for AI\end{tabular}                 & 2019 & IA Latam                                                                                                                               & Ind                                                       & Chile          & \yes      & \yes            & \bigyes   & \yes          \\ 
	
	\bottomrule
	
	\label{tbl:frameworks}
\end{longtable}

%% file: main.bbl
\begin{thebibliography}{10}

\bibitem{MLSecurity3}
Martin Abadi, Andy Chu, Ian Goodfellow, H~Brendan McMahan, Ilya Mironov, Kunal
  Talwar, and Li~Zhang.
\newblock {Deep learning with differential privacy}.
\newblock In {\em Proceedings of the 2016 ACM SIGSAC Conference on Computer and
  Communications Security}, pages 308--318. ACM, 2016.

\bibitem{lineage6}
Eleanor Ainy, Pierre Bourhis, Susan~B Davidson, Daniel Deutch, and Tova Milo.
\newblock {Approximated summarization of data provenance}.
\newblock In {\em Proceedings of the 24th ACM International on Conference on
  Information and Knowledge Management}, pages 483--492. ACM, 2015.

\bibitem{aitken2016}
Mhairi Aitken, Sarah Cunningham-Burley, and Claudia Pagliari.
\newblock {Moving from trust to trustworthiness: Experiences of public
  engagement in the Scottish Health Informatics Programme}.
\newblock {\em Science and Public Policy}, 43(5):713--723, 2016.

\bibitem{altman1992introduction}
Naomi~S Altman.
\newblock {An introduction to kernel and nearest-neighbor nonparametric
  regression}.
\newblock {\em The American Statistician}, 46(3):175--185, 1992.

\bibitem{andras2018}
Peter Andras, Lukas Esterle, Michael Guckert, The~Anh Han, Peter~R Lewis,
  Kristina Milanovic, Terry Payne, Cedric Perret, Jeremy Pitt, Simon~T Powers,
  and {others}.
\newblock {Trusting Intelligent Machines: Deepening Trust Within
  Socio-Technical Systems}.
\newblock {\em IEEE Technology and Society Magazine}, 37(4):76--83, 2018.

\bibitem{athey2015machine}
Susan Athey and Guido~W Imbens.
\newblock {Machine learning methods for estimating heterogeneous causal
  effects}.
\newblock {\em stat}, 1050(5):1--26, 2015.

\bibitem{avizienis04}
A~Avizienis, J~. Laprie, B~Randell, and C~Landwehr.
\newblock {Basic concepts and taxonomy of dependable and secure computing}.
\newblock {\em IEEE Transactions on Dependable and Secure Computing},
  1(1):11--33, 1 2004.

\bibitem{lineage9}
Akanksha Baid, Wentao Wu, Chong Sun, AnHai Doan, and Jeffrey~F Naughton.
\newblock {On Debugging Non-Answers in Keyword Search Systems.}
\newblock In {\em EDBT}, pages 37--48, 2015.

\bibitem{fintechFairness}
Robert Bartlett, Adair Morse, Richard Stanton, and Nancy Wallace.
\newblock {Consumer-Lending Discrimination in the Era of FinTech}.
\newblock {\em Unpublished working paper. University of California, Berkeley},
  2018.

\bibitem{lineage10}
Nicole Bidoit, Melanie Herschel, and Katerina Tzompanaki.
\newblock {Query-based why-not provenance with nedexplain}.
\newblock In {\em Extending database technology (EDBT)}, 2014.

\bibitem{trainingAttack4}
Battista Biggio, Igino Corona, Giorgio Fumera, Giorgio Giacinto, and Fabio
  Roli.
\newblock {Bagging classifiers for fighting poisoning attacks in adversarial
  classification tasks}.
\newblock In {\em International workshop on multiple classifier systems}, pages
  350--359. Springer, 2011.

\bibitem{quantitativeSecurity2}
Battista Biggio, Igino Corona, Blaine Nelson, Benjamin I~P Rubinstein, Davide
  Maiorca, Giorgio Fumera, Giorgio Giacinto, and Fabio Roli.
\newblock {Security evaluation of support vector machines in adversarial
  environments}.
\newblock In {\em Support Vector Machines Applications}, pages 105--153.
  Springer, 2014.

\bibitem{trainingAttack3}
Battista Biggio, Giorgio Fumera, and Fabio Roli.
\newblock {Multiple classifier systems for robust classifier design in
  adversarial environments}.
\newblock {\em International Journal of Machine Learning and Cybernetics},
  1(1-4):27--41, 2010.

\bibitem{securityassessment}
Battista Biggio, Giorgio Fumera, and Fabio Roli.
\newblock {Security evaluation of pattern classifiers under attack}.
\newblock {\em IEEE transactions on knowledge and data engineering},
  26(4):984--996, 2014.

\bibitem{biggio2011support}
Battista Biggio, Blaine Nelson, and Pavel Laskov.
\newblock {Support vector machines under adversarial label noise}.
\newblock In {\em Asian Conference on Machine Learning}, pages 97--112, 2011.

\bibitem{bishop2007penetration}
Matt Bishop.
\newblock {About penetration testing}.
\newblock {\em IEEE Security {\&} Privacy}, 5(6):84--87, 2007.

\bibitem{Blum1997}
Avrim~L Blum and Pat Langley.
\newblock {Selection of relevant features and examples in machine learning}.
\newblock {\em Artificial Intelligence}, 97(1-2):245--271, 1997.

\bibitem{lineage1}
Peter Buneman, Adriane Chapman, and James Cheney.
\newblock {Provenance management in curated databases}.
\newblock In {\em Proceedings of the 2006 ACM SIGMOD international conference
  on Management of data}, pages 539--550. ACM, 2006.

\bibitem{Cai2018}
Jie Cai, Jiawei Luo, Shulin Wang, and Sheng Yang.
\newblock {Feature selection in machine learning: A new perspective}.
\newblock {\em Neurocomputing}, 300:70--79, 2018.

\bibitem{lineage2}
Adriane~P Chapman, Hosagrahar~V Jagadish, and Prakash Ramanan.
\newblock {Efficient provenance storage}.
\newblock In {\em Proceedings of the 2008 ACM SIGMOD international conference
  on Management of data}, pages 993--1006. ACM, 2008.

\bibitem{lineage3}
James Cheney, Amal Ahmed, and Umut~A Acar.
\newblock {Provenance as dependency analysis}.
\newblock In {\em International Symposium on Database Programming Languages},
  pages 138--152. Springer, 2007.

\bibitem{chin2012measuring}
Erika Chin, Adrienne~Porter Felt, Vyas Sekar, and David Wagner.
\newblock {Measuring user confidence in smartphone security and privacy}.
\newblock In {\em Proceedings of the eighth symposium on usable privacy and
  security}, page~1. ACM, 2012.

\bibitem{corbett2018measure}
Sam Corbett-Davies and Sharad Goel.
\newblock {The Measure and Mismeasure of Fairness: A Critical Review of Fair
  Machine Learning}.
\newblock {\em arXiv preprint arXiv:1808.00023}, 2018.

\bibitem{cunningham2006}
Sarah Cunningham-Burley.
\newblock {Public knowledge and public trust}.
\newblock {\em Public Health Genomics}, 9(3):204--210, 2006.

\bibitem{dataScienceView}
Brian d'Alessandro, Cathy O'Neil, and Tom LaGatta.
\newblock {Conscientious classification: A data scientist's guide to
  discrimination-aware classification}.
\newblock {\em Big data}, 5(2):120--134, 2017.

\bibitem{demontis2017yes}
Ambra Demontis, Marco Melis, Battista Biggio, Davide Maiorca, Daniel Arp,
  Konrad Rieck, Igino Corona, Giorgio Giacinto, and Fabio Roli.
\newblock {Yes, machine learning can be more secure! a case study on android
  malware detection}.
\newblock {\em IEEE Transactions on Dependable and Secure Computing}, 2017.

\bibitem{lineage11}
Daniel Deutch, Nave Frost, and Amir Gilad.
\newblock {Nlprov: Natural language provenance}.
\newblock {\em Proceedings of the VLDB Endowment}, 9(13):1537--1540, 2016.

\bibitem{lineage7}
Daniel Deutch, Amir Gilad, and Yuval Moskovitch.
\newblock {Selective provenance for datalog programs using top-k queries}.
\newblock {\em Proceedings of the VLDB Endowment}, 8(12):1394--1405, 2015.

\bibitem{dietz2006}
Graham Dietz and Deanne~N Den~Hartog.
\newblock {Measuring trust inside organisations}.
\newblock {\em Personnel review}, 35(5):557--588, 2006.

\bibitem{dietz2012}
Graham Dietz and Nicole Gillespie.
\newblock {\em {Recovery of Trust: Case Studies of Organisational Failures and
  Trust Repair}}, volume~5.
\newblock Institute of Business Ethics London, 2012.

\bibitem{doshi2017towards}
Finale Doshi-Velez and Been Kim.
\newblock {Towards a rigorous science of interpretable machine learning}.
\newblock {\em arXiv preprint arXiv:1702.08608}, 2017.

\bibitem{drucker1999support}
Harris Drucker, Donghui Wu, and Vladimir~N Vapnik.
\newblock {Support vector machines for spam categorization}.
\newblock {\em IEEE Transactions on Neural networks}, 10(5):1048--1054, 1999.

\bibitem{duwairi2014study}
Rehab Duwairi and Mahmoud El-Orfali.
\newblock {A study of the effects of preprocessing strategies on sentiment
  analysis for Arabic text}.
\newblock {\em Journal of Information Science}, 40(4):501--513, 2014.

\bibitem{dwork2011differential}
Cynthia Dwork.
\newblock {Differential privacy}.
\newblock {\em Encyclopedia of Cryptography and Security}, pages 338--340,
  2011.

\bibitem{fairnessAwareness}
Cynthia Dwork, Moritz Hardt, Toniann Pitassi, Omer Reingold, and Richard Zemel.
\newblock {Fairness through awareness}.
\newblock In {\em Proceedings of the 3rd innovations in theoretical computer
  science conference}, pages 214--226. ACM, 2012.

\bibitem{erhan2010does}
Dumitru Erhan, Yoshua Bengio, Aaron Courville, Pierre-Antoine Manzagol, Pascal
  Vincent, and Samy Bengio.
\newblock {Why does unsupervised pre-training help deep learning?}
\newblock {\em Journal of Machine Learning Research}, 11(Feb):625--660, 2010.

\bibitem{demographic}
Michael Feldman, Sorelle~A Friedler, John Moeller, Carlos Scheidegger, and
  Suresh Venkatasubramanian.
\newblock {Certifying and removing disparate impact}.
\newblock In {\em Proceedings of the 21th ACM SIGKDD International Conference
  on Knowledge Discovery and Data Mining}, pages 259--268. ACM, 2015.

\bibitem{ethicalGraph}
Jessica Fjeld, Hannah Hilligoss, Nele Achten, Maia~Levy Daniel, Sally Kagay,
  and Joshua Feldman.
\newblock {Principled Artificial Intelligence: Mapping Consensus and Divergence
  in Ethical and Rights-Based Approaches}.
\newblock 2019.

\bibitem{ghodsi2017safetynets}
Zahra Ghodsi, Tianyu Gu, and Siddharth Garg.
\newblock {Safetynets: Verifiable execution of deep neural networks on an
  untrusted cloud}.
\newblock In {\em Advances in Neural Information Processing Systems}, pages
  4672--4681, 2017.

\bibitem{gillespie2009}
Nicole Gillespie and Graham Dietz.
\newblock {Trust repair after an organization-level failure}.
\newblock {\em Academy of Management Review}, 34(1):127--145, 2009.

\bibitem{Goncalves2010}
Carlos~Adriano Gon{\c{c}}alves, Celia~Talma Gon{\c{c}}alves, Rui Camacho, and
  Eugenio~C Oliveira.
\newblock {The impact of Pre-Processing on the Classification of MEDLINE
  Documents}.
\newblock {\em Pattern Recognition in Information Systems, Proceedings of the
  10th International Workshop on Pattern Recognition in Information Systems,
  PRIS 2010, In conjunction with ICEIS 2010}, page~10, 2010.

\bibitem{zelaya2019volatility}
Carlos~Vladimiro Gonz{\'{a}}lez~Zelaya.
\newblock {Towards Explaining the Effects of Data Preprocessing on Machine
  Learning}.
\newblock {\em 2019 IEEE 35th International Conference on Data Engineering
  (ICDE)}, 2019.

\bibitem{explainableSurvey}
Riccardo Guidotti, Anna Monreale, Salvatore Ruggieri, Franco Turini, Fosca
  Giannotti, and Dino Pedreschi.
\newblock {A survey of methods for explaining black box models}.
\newblock {\em ACM computing surveys (CSUR)}, 51(5):93, 2018.

\bibitem{lineageSurvey}
Himanshu Gupta, Sameep Mehta, Sandeep Hans, Bapi Chatterjee, Pranay Lohia, and
  C~Rajmohan.
\newblock {Provenance in context of Hadoop as a Service (HaaS)-State of the Art
  and Research Directions}.
\newblock In {\em 2017 IEEE 10th International Conference on Cloud Computing
  (CLOUD)}, pages 680--683. IEEE, 2017.

\bibitem{hagendijk2006}
Rob Hagendijk and Alan Irwin.
\newblock {Public deliberation and governance: engaging with science and
  technology in contemporary Europe}.
\newblock {\em Minerva}, 44(2):167--184, 2006.

\bibitem{equalizedOdds}
Moritz Hardt, Eric Price, Nati Srebro, and {others}.
\newblock {Equality of opportunity in supervised learning}.
\newblock In {\em Advances in neural information processing systems}, pages
  3315--3323, 2016.

\bibitem{hu2017generating}
Weiwei Hu and Ying Tan.
\newblock {Generating adversarial malware examples for black-box attacks based
  on GAN}.
\newblock {\em arXiv preprint arXiv:1702.05983}, 2017.

\bibitem{jain1999data}
Anil~K Jain, M~Narasimha Murty, and Patrick~J Flynn.
\newblock {Data clustering: a review}.
\newblock {\em ACM computing surveys (CSUR)}, 31(3):264--323, 1999.

\bibitem{Joseph2016a}
Matthew Joseph, Michael Kearns, Jamie Morgenstern, Seth Neel, and Aaron Roth.
\newblock {Rawlsian Fairness for Machine Learning}.
\newblock {\em FATML}, pages 1--26, 2016.

\bibitem{fairclassifier}
Toshihiro Kamishima, Shotaro Akaho, Hideki Asoh, and Jun Sakuma.
\newblock {Fairness-aware classifier with prejudice remover regularizer}.
\newblock In {\em Joint European Conference on Machine Learning and Knowledge
  Discovery in Databases}, pages 35--50. Springer, 2012.

\bibitem{kim2015ibcm}
Been Kim, Elena Glassman, Brittney Johnson, and Julie Shah.
\newblock {iBCM: Interactive Bayesian case model empowering humans via
  intuitive interaction}.
\newblock 2015.

\bibitem{koszegi2019high}
Sabine~Theresia Koszegi.
\newblock {High-Level Expert Group on Artificial Intelligence}, 2019.

\bibitem{krizhevsky2009learning}
Alex Krizhevsky, Geoffrey Hinton, and {others}.
\newblock {Learning multiple layers of features from tiny images}.
\newblock Technical report, Citeseer, 2009.

\bibitem{krizhevsky2012imagenet}
Alex Krizhevsky, Ilya Sutskever, and Geoffrey~E Hinton.
\newblock {Imagenet classification with deep convolutional neural networks}.
\newblock In {\em Advances in neural information processing systems}, pages
  1097--1105, 2012.

\bibitem{counterfactual}
Matt~J Kusner, Joshua Loftus, Chris Russell, and Ricardo Silva.
\newblock {Counterfactual fairness}.
\newblock In {\em Advances in Neural Information Processing Systems}, pages
  4066--4076, 2017.

\bibitem{trainingAttack6}
Ricky Laishram and Vir~Virander Phoha.
\newblock {Curie: A method for protecting SVM Classifier from Poisoning
  Attack}.
\newblock {\em arXiv preprint arXiv:1606.01584}, 2016.

\bibitem{Lakkaraju2017}
Himabindu Lakkaraju, Ece Kamar, Rich Caruana, and Jure Leskovec.
\newblock {Interpretable {\&} Explorable Approximations of Black Box Models}.
\newblock 7 2017.

\bibitem{quantitativeSecurity1}
Pavel Laskov and Marius Kloft.
\newblock {A framework for quantitative security analysis of machine learning}.
\newblock In {\em Proceedings of the 2nd ACM workshop on Security and
  artificial intelligence}, pages 1--4. ACM, 2009.

\bibitem{socialTrustTechnology}
Xin Li, Traci~J Hess, and Joseph~S Valacich.
\newblock {Why do we trust new technology? A study of initial trust formation
  with organizational information systems}.
\newblock {\em The Journal of Strategic Information Systems}, 17(1):39--71,
  2008.

\bibitem{lipton2016mythos}
Zachary~C Lipton.
\newblock {The mythos of model interpretability}.
\newblock {\em arXiv preprint arXiv:1606.03490}, 2016.

\bibitem{securitySurvey}
Qiang Liu, Pan Li, Wentao Zhao, Wei Cai, Shui Yu, and Victor C~M Leung.
\newblock {A survey on security threats and defensive techniques of machine
  learning: A data driven view}.
\newblock {\em IEEE access}, 6:12103--12117, 2018.

\bibitem{Lou2013}
Yin Lou, Rich Caruana, Johannes Gehrke, and Giles Hooker.
\newblock {Accurate intelligible models with pairwise interactions}.
\newblock {\em Proceedings of the 19th ACM SIGKDD international conference on
  Knowledge discovery and data mining - KDD '13}, page 623, 2013.

\bibitem{statiticalFramework}
Kristian Lum and James Johndrow.
\newblock {A statistical framework for fair predictive algorithms}.
\newblock {\em arXiv preprint arXiv:1610.08077}, 2016.

\bibitem{lundberg2017unified}
Scott~M Lundberg and Su-In Lee.
\newblock {A unified approach to interpreting model predictions}.
\newblock In {\em Advances in Neural Information Processing Systems}, pages
  4765--4774, 2017.

\bibitem{luong2011k}
Binh~Thanh Luong, Salvatore Ruggieri, and Franco Turini.
\newblock {k-NN as an implementation of situation testing for discrimination
  discovery and prevention}.
\newblock In {\em Proceedings of the 17th ACM SIGKDD international conference
  on Knowledge discovery and data mining}, pages 502--510. ACM, 2011.

\bibitem{lineage8}
Peter Macko, Daniel Margo, and Margo Seltzer.
\newblock {Local clustering in provenance graphs}.
\newblock In {\em Proceedings of the 22nd ACM international conference on
  Information {\&} Knowledge Management}, pages 835--840. ACM, 2013.

\bibitem{mayer1995}
Roger~C Mayer, James~H Davis, and F~David Schoorman.
\newblock {An integrative model of organizational trust}.
\newblock {\em Academy of management review}, 20(3):709--734, 1995.

\bibitem{blackBoxMachineLearning}
Donald Michie, David~J Spiegelhalter, C~C Taylor, and {others}.
\newblock {Machine learning}.
\newblock {\em Neural and Statistical Classification}, 13, 1994.

\bibitem{montavon2018methods}
Grégoire Montavon, Wojciech Samek, and Klaus-Robert M{\"{u}}ller.
\newblock {Methods for interpreting and understanding deep neural networks}.
\newblock {\em Digital Signal Processing}, 73:1--15, 2018.

\bibitem{narayanan2018translation}
Arvind Narayanan.
\newblock {Translation tutorial: 21 fairness definitions and their politics}.
\newblock In {\em Proc. Conf. Fairness Accountability Transp., New York, USA},
  2018.

\bibitem{secureCompuation}
Olga Ohrimenko, Felix Schuster, Cédric Fournet, Aastha Mehta, Sebastian
  Nowozin, Kapil Vaswani, and Manuel Costa.
\newblock {Oblivious multi-party machine learning on trusted processors}.
\newblock In {\em 25th {\$}{\{}{\$}USENIX{\$}{\}}{\$} Security Symposium
  ({\$}{\{}{\$}USENIX{\$}{\}}{\$} Security 16)}, pages 619--636, 2016.

\bibitem{papernot2016towards}
Nicolas Papernot, Patrick McDaniel, Arunesh Sinha, and Michael Wellman.
\newblock {Towards the science of security and privacy in machine learning}.
\newblock {\em arXiv preprint arXiv:1611.03814}, 2016.

\bibitem{MLSecuritySurvey}
Nicolas Papernot, Patrick McDaniel, Arunesh Sinha, and Michael~P Wellman.
\newblock {SoK: Security and privacy in machine learning}.
\newblock In {\em 2018 IEEE European Symposium on Security and Privacy
  (EuroS{\&}P)}, pages 399--414. IEEE, 2018.

\bibitem{lineage4}
Christopher R{\'{e}} and Dan Suciu.
\newblock {Approximate lineage for probabilistic databases}.
\newblock {\em Proceedings of the VLDB Endowment}, 1(1):797--808, 2008.

\bibitem{lime}
Marco~Tulio Ribeiro, Sameer Singh, and Carlos Guestrin.
\newblock {Why should i trust you?: Explaining the predictions of any
  classifier}.
\newblock In {\em Proceedings of the 22nd ACM SIGKDD international conference
  on knowledge discovery and data mining}, pages 1135--1144. ACM, 2016.

\bibitem{rivest1987learning}
Ronald~L Rivest.
\newblock {Learning decision lists}.
\newblock {\em Machine learning}, 2(3):229--246, 1987.

\bibitem{interdisciplinary}
Andrea Romei and Salvatore Ruggieri.
\newblock {A multidisciplinary survey on discrimination analysis}.
\newblock {\em The Knowledge Engineering Review}, 29(5):582--638, 2014.

\bibitem{trainingAttack2}
Benjamin I~P Rubinstein, Blaine Nelson, Ling Huang, Anthony~D Joseph, Shing-hon
  Lau, Satish Rao, Nina Taft, and J~Doug Tygar.
\newblock {Antidote: understanding and defending against poisoning of anomaly
  detectors}.
\newblock In {\em Proceedings of the 9th ACM SIGCOMM conference on Internet
  measurement}, pages 1--14. ACM, 2009.

\bibitem{MPCCrypto}
Reza Shokri and Vitaly Shmatikov.
\newblock {Privacy-preserving deep learning}.
\newblock In {\em Proceedings of the 22nd ACM SIGSAC conference on computer and
  communications security}, pages 1310--1321. ACM, 2015.

\bibitem{shrikumar2017learning}
Avanti Shrikumar, Peyton Greenside, and Anshul Kundaje.
\newblock {Learning important features through propagating activation
  differences}.
\newblock In {\em Proceedings of the 34th International Conference on Machine
  Learning-Volume 70}, pages 3145--3153. JMLR. org, 2017.

\bibitem{trustMLSocial}
Keng Siau and Weiyu Wang.
\newblock {Building trust in artificial intelligence, machine learning, and
  robotics}.
\newblock {\em Cutter Business Technology Journal}, 31(2):47--53, 2018.

\bibitem{singh2018decision}
Jatinder Singh, Jennifer Cobbe, and Chris Norval.
\newblock {Decision Provenance: Harnessing data flow for accountable systems}.
\newblock {\em IEEE Access}, 7:6562--6574, 2018.

\bibitem{sutskever2014sequence}
Ilya Sutskever, Oriol Vinyals, and Quoc~V Le.
\newblock {Sequence to sequence learning with neural networks}.
\newblock In {\em Advances in neural information processing systems}, pages
  3104--3112, 2014.

\bibitem{sutton2018reinforcement}
Richard~S Sutton and Andrew~G Barto.
\newblock {\em {Reinforcement learning: An introduction}}.
\newblock MIT press, 2018.

\bibitem{szegedy2013intriguing}
Christian Szegedy, Wojciech Zaremba, Ilya Sutskever, Joan Bruna, Dumitru Erhan,
  Ian Goodfellow, and Rob Fergus.
\newblock {Intriguing properties of neural networks}.
\newblock {\em arXiv preprint arXiv:1312.6199}, 2013.

\bibitem{Uysal2014}
Alper~Kursat Uysal and Serkan Gunal.
\newblock {The impact of preprocessing on text classification}.
\newblock {\em Information Processing and Management}, 50(1):104--112, 2014.

\bibitem{walfish2015verifying}
Michael Walfish and Andrew~J Blumberg.
\newblock {Verifying computations without reexecuting them}.
\newblock {\em Communications of the ACM}, 58(2):74--84, 2015.

\bibitem{whittlestone2019role}
Jess Whittlestone, Rune Nyrup, Anna Alexandrova, and Stephen Cave.
\newblock {The Role and Limits of Principles in AI Ethics: Towards a Focus on
  Tensions}.
\newblock In {\em Proceedings of the AAAI/ACM Conference on AI Ethics and
  Society, Honolulu, HI, USA}, pages 27--28, 2019.

\bibitem{wood2003thin}
Simon~N Wood.
\newblock {Thin plate regression splines}.
\newblock {\em Journal of the Royal Statistical Society: Series B (Statistical
  Methodology)}, 65(1):95--114, 2003.

\bibitem{lineage12}
Eugene Wu, Samuel Madden, and Michael Stonebraker.
\newblock {Subzero: a fine-grained lineage system for scientific databases}.
\newblock In {\em 2013 IEEE 29th International Conference on Data Engineering
  (ICDE)}, pages 865--876. IEEE, 2013.

\bibitem{wynne1992misunderstood}
Brian Wynne.
\newblock Misunderstood misunderstanding: social identities and public uptake
  of science.
\newblock {\em Public understanding of science}, 1(3):281--304, 1992.

\bibitem{wynne1996reflexive}
Brian Wynne.
\newblock {A reflexive view of the expert-lay knowledge divide}.
\newblock {\em Risk, environment and modernity: Towards a new ecology}, 40:44,
  1996.

\bibitem{wynne2006}
Brian Wynne.
\newblock {Public engagement as a means of restoring public trust in
  science--hitting the notes, but missing the music?}
\newblock {\em Public Health Genomics}, 9(3):211--220, 2006.

\bibitem{lineage5}
Yulai Xie, Kiran-Kumar Muniswamy-Reddy, Dan Feng, Yan Li, and Darrell D~E Long.
\newblock {Evaluation of a hybrid approach for efficient provenance storage}.
\newblock {\em ACM Transactions on Storage (TOS)}, 9(4):14, 2013.

\bibitem{yuki2005cross}
Masaki Yuki, William~W Maddux, Marilynn~B Brewer, and Kosuke Takemura.
\newblock {Cross-cultural differences in relationship-and group-based trust}.
\newblock {\em Personality and Social Psychology Bulletin}, 31(1):48--62, 2005.

\bibitem{zafar2019discrimination}
Muhammad~Bilal Zafar.
\newblock {Discrimination in Algorithmic Decision Making: From Principles to
  Measures and Mechanisms}.
\newblock 2019.

\bibitem{disparateOdds2}
Muhammad~Bilal Zafar, Isabel Valera, Manuel Gomez~Rodriguez, and Krishna~P
  Gummadi.
\newblock {Fairness beyond disparate treatment {\&} disparate impact: Learning
  classification without disparate mistreatment}.
\newblock In {\em Proceedings of the 26th International Conference on World
  Wide Web}, pages 1171--1180. International World Wide Web Conferences
  Steering Committee, 2017.

\bibitem{fairnessOverview}
Indr\.e {\v{Z}}liobait{\textbackslash}.e.
\newblock {Measuring discrimination in algorithmic decision making}.
\newblock {\em Data Mining and Knowledge Discovery}, 31(4):1060--1089, 2017.

\end{thebibliography}
